\documentclass[pre,twocolumn,floatfix]{revtex4-1}

\usepackage{amsmath}
\usepackage{amssymb,amsfonts,latexsym}
\usepackage{bm}
\usepackage[mathcal]{euscript}
\usepackage{graphicx}
\usepackage{epsfig}
\usepackage{grfext}

\usepackage{natbib}
\usepackage{url}
\usepackage{bm}

\usepackage{amscd}
\usepackage{epstopdf}
\usepackage{verbatim}
\usepackage[usenames]{color}
\usepackage{array}
\usepackage{xspace}
\usepackage{xstring}
\usepackage[normalem]{ulem}
\usepackage{xr}
\usepackage[breaklinks=true,
            pdfborder={0 0 1},
            colorlinks=true,
            linkcolor=black,
            citecolor=blue,
            urlcolor=blue]{hyperref}

\PrependGraphicsExtensions*{.pdf,.PDF}  

\newcommand{\eg}{\textit{e.g.}\xspace}

\begin{document}
\title{Recent advances in coarse-grained modeling of virus assembly}
\author{Michael F. Hagan}\email{hagan@brandeis.edu}
\affiliation{Martin Fisher School of Physics, Brandeis University, Waltham, MA 02453, USA}
\author{Roya Zandi}\email{royaz@ucr.edu}
\affiliation{Department of Physics and Astronomy, University of California, Riverside, California 92521, USA}

\begin{abstract}
In many virus families, tens to thousands of proteins assemble spontaneously into a capsid (protein shell) while packaging the genomic nucleic acid. This review summarizes recent advances in computational modeling of these dynamical processes. We present an overview of recent technological and algorithmic developments, which are enabling simulations to describe the large ranges of length-and time-scales relevant to assembly, under conditions more closely matched to experiments than in earlier work. We then describe two examples in which computational modeling has recently provided an important complement to experiments.
\end{abstract}


\maketitle

Capsid assembly and packaging of the genome are essential steps in the formation of an infectious virus. Thus, elucidating the mechanisms by which assembly proceeds could identify important targets for antiviral drugs and would advance our fundamental understanding of the viral lifecycle. However, assembly and packaging pathways remain incompletely understood for many viruses because most intermediates are transient and therefore undetectable or characterized only with low resolution. Computer simulations of virus assembly have overcome this limitation by revealing features of assembly processes that are not accessible to experiments alone. However, the large size of a virus (15-1000 nm) and the timescales required for assembly (ms-hours) prohibit simulating capsid formation with atomic-resolution, except for specific steps \cite{Jiang2015}. To this end, researchers have relied on simplified models, which aim to coarse-grain over atomic-scale details while accurately describing the essential physical features that control assembly.

This review describes recent advances in coarse-grained models of capsid assembly. We begin with a brief overview of models and simulation methodologies, followed by recent applications of these approaches. To accommodate space limitations, we limit our discussion of applications to two areas which have recently been the subject of intense modeling activity: the role of nucleic acids in the assembly of icosahedral viruses, and assembly of the mature HIV capsid.

\begin{figure*} [hbt]
\centering{\includegraphics[width=0.75\textwidth]{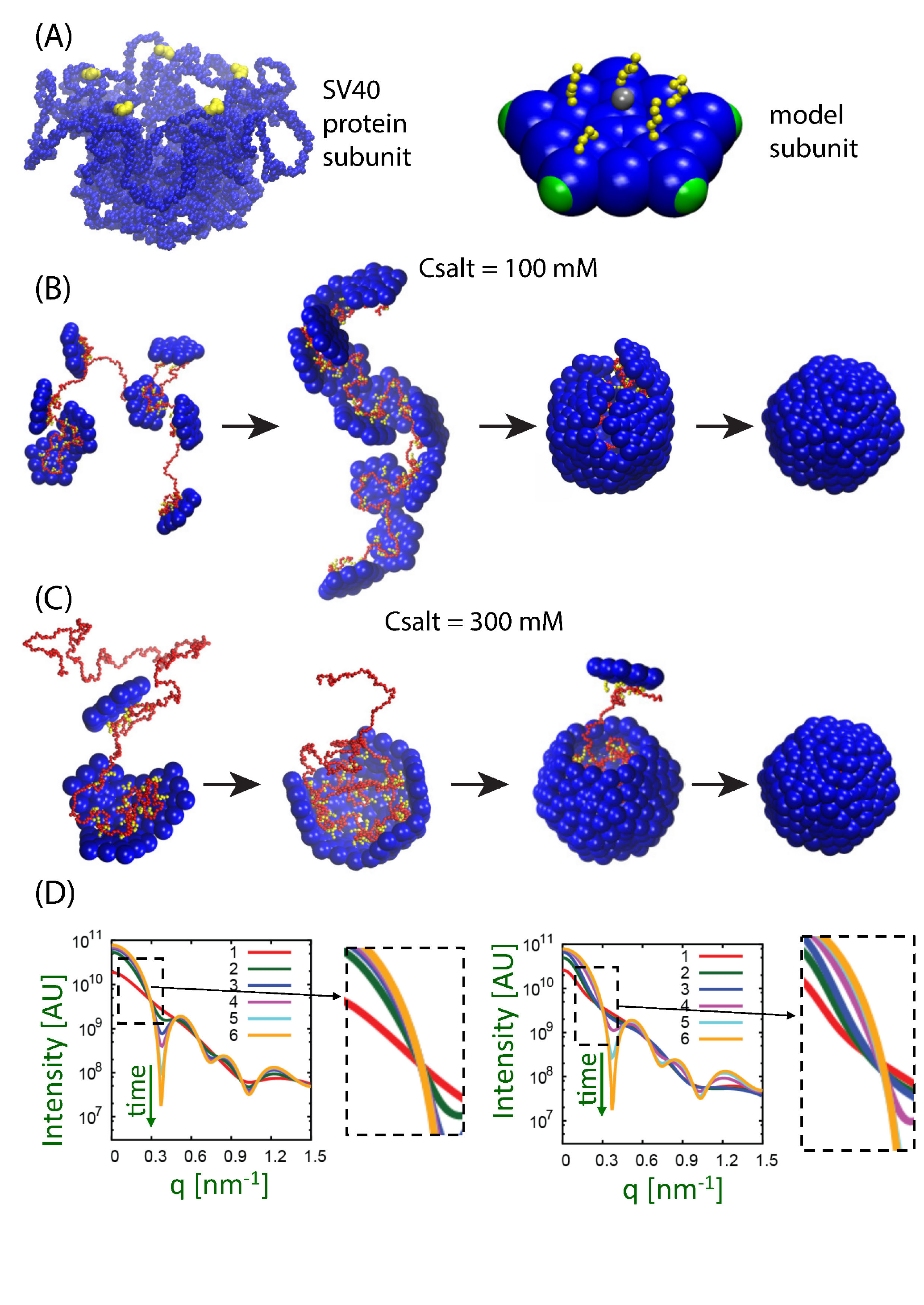}}
\caption{A particle-based model for capsid assembly around a linear polymer. {\bf (A)} (Left) Image of the crystal structure of a homopentamer of the SV40 capsid protein, the elemental subunit for SV40 capsid assembly, with visible portions of RNA binding domains in yellow. (Right) Image of a coarse-grained model subunit, with RNA binding domains shown in yellow. {\bf (B),(C)} Snapshots from typical simulation trajectories illustrating two classes of pathways for assembly around a polymer or RNA. In (B), strong protein-RNA interactions lead to an `en masse' mechanism, in which proteins rapidly adsorb onto the RNA in a disordered manner, followed by cooperative rearrangements to form an assembled capsid. In (C), weaker protein-RNA interactions drive a nucleation-and-growth mechanism, in which a small, ordered nucleus forms, followed by sequential addition of protein subunits. {\bf (D)} Time-resolved small angle x-ray scattering (SAXS) profiles estimated from simulation trajectories corresponding to the (left) nucleation-and-growth mechanism and (right) en masse mechanism. The zoom-in to the right of each plot illustrates one of the distinguishing features of the SAXS profiles --- the nucleation-and-growth mechanism leads to an isosbestic point among profiles measured at different times, whereas the en masse mechanism does not (at early times). Figure adapted from Ref.~\cite{Perlmutter2014}.}
\label{fig:JasonFigure}
\end{figure*}

\section*{Coarse-grained models for capsid assembly}
One approach to model development seeks to describe a specific physical system with the greatest accuracy allowed by computational constraints, and by systematically coarse-graining from atomistic simulations  (\eg \cite{Davtyan2015,Dama2013}). However, the conformational dynamics of capsid proteins restricts the accuracy of such techniques, and the complexity of the resulting coarse-grained models has limited their application to assembly. Therefore, capsid assembly models have relied on a combination of atomistic simulations, structural data, and fitting model parameters to kinetics and thermodynamic data. Often, the aim has been to construct the simplest model consistent with experimental data, to discover general, fundamental insights about capsid assembly.

Models for virus assembly can be separated into three classes. In the first,
 the time evolution of concentrations of capsid intermediates is represented by a system of rate equations \cite{Zlotnick1999,Endres2002}. Formulation of the model requires specifying the state space (the set of all possible assembly intermediates) and transition rates between each pair of intermediates.  The rate equations can be numerically integrated \cite{Zlotnick1999,Endres2002,Moisant2010,Zandi2006,Schoot2007}, or trajectories consistent with the rate equations can be stochastically sampled using Gillespie-type algorithms \cite{Gillespie1977,Dykeman2013a,Zhang2006a,Xie2012,Smith2014,Dykeman2014}, and transition rates can be fit against experimental data \cite{Kumar2010,Xie2012,Zlotnick1999,Tsiang2012}. Despite their simplicity, such models reproduce many experimental observations on capsid assembly.

In the next class of models (particle-based simulations),  subunits interact through pair potentials that drive assembly toward an ordered low-energy structure (\eg an icosahedral shell \cite{Schwartz1998,Hagan2006,Nguyen2007,Rapaport1999,Rapaport2012,Baschek2012,Perlmutter2014,Perlmutter2015b,Boettcher2015}, Fig.~\ref{fig:JasonFigure}). Subunit motions are explicitly tracked by numerically integrating equations of motion (\eg using molecular dynamics, Brownian dynamics (BD), or discontinuous molecular dynamics \cite{Rapaport2012,Perlmutter2014,Nguyen2007}.
 The third approach combines aspects of Gillespie-type and particle-based models, modeling assembly through the irreversible addition of triangular subunits to growing edges of an incomplete shell \cite{Hicks2006,Levandovsky2009,Wagner2015a}. The shell is treated as an elastic sheet that relaxes to its minimum energy configuration after each accretion.

Of these approaches, particle-based simulations enable (at least in principle) the fewest assumptions about intermediate geometries and the highest resolution description of proteins. However, their high computational cost has limited model resolution, simulation timescales, and system sizes \cite{Hagan2014,Perilla2015}.

Recent algorithmic advances are beginning to overcome this limitation. For example, Grime and Voth \cite{Grime2014} designed an efficient parallelization scheme for spatially heterogeneous particle concentrations (which occur during assembly simulations with implicit solvent). Algorithms performing rigid body dynamics on GPUs show significant speedup in comparison to conventional CPUs. For example, the package HOOMD \cite{Nguyen2011} has been used to simulate virus assembly around nucleic acids and on membranes \cite{Perlmutter2013,Perlmutter2014,Perlmutter2015b,Ruiz-Herrero2015}. Zuckerman and coworkers \cite{Spiriti2015} have shown that calculation of interparticle potentials between rigid bodies containing many interaction sites (such as high resolution models of proteins) can be speeded up by tabulation in distance and orientation space.
Finally, enhanced sampling methods can focus computational time on critical but rare events, such as crossing nucleation barriers. Methods well-suited for the diverse pathways typical of assembly systems include multiple state transition path sampling \cite{Newton2015}, Markov state models \cite{Perkett2014}, weighted ensemble dynamics \cite{Spiriti2015}, and diffusion maps \cite{Long2014}.

\begin{figure*} [hbt]
\centering{\includegraphics[width=0.7\textwidth]{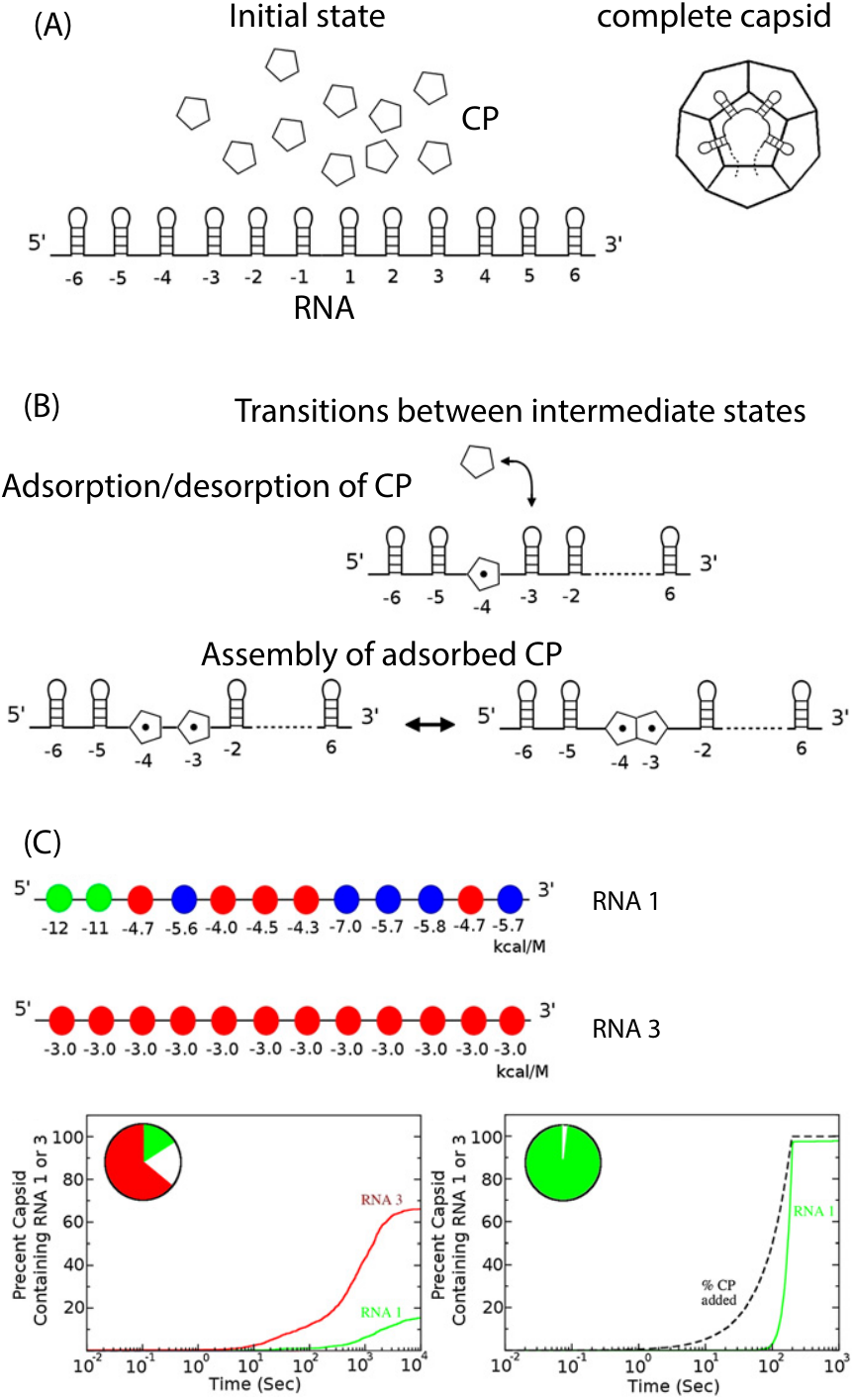}}
\caption{A Gillespie-type model for packaging signal (PS) mediated assembly around RNA. {\bf (A)} The initial and final states of the model, corresponding respectively to a naked RNA molecule with unassociated capsid proteins, and a dodecahedral model capsid assembled around  RNA. Potential high affinity binding sites (PSs) are denoted as hairpins. {\bf (B)} Examples of intermediate states and transitions between them, corresponding to the adsorption/desorption of a capsid protein to/from the RNA, and association/dissociation of two capsid proteins adsorbed on the RNA. (C) Simulated competition for packaging by capsid proteins of 2000 RNA molecules containing PSs (RNA1) against 60,000 cellular RNAs (no PSs), under (left) constant total protein concentration or (right) a steadily increasing protein concentration. Figures adapted from Ref.~\cite{Dykeman2014}.}
\label{fig:DykemanFigure}
\end{figure*}

\section*{Applications of coarse-grained modeling} 
There have been extensive applications of coarse-grained models to understand the assembly of small icosahedral capsids (reviewed in \cite{Hagan2014}). Initial studies focused on modeling in vitro experiments in which pure proteins assemble into empty capsids. Recent works have included effects of crowding \cite{Smith2014}, assembly on membranes \cite{Matthews2013a,Ruiz-Herrero2015}, and how assembly changes when subunits are generated during the course of a reaction, either by protein translation or advective transport \cite{Hagan2011,Boettcher2015,Castelnovo2014}. As noted above, we focus here on the role of nucleic acids (NAs) and other negatively charged cargos in the assembly of icosahedral viruses, and assembly of the mature HIV shell.

{\it Assembly around nucleic acids and other cargoes.}
While there are several mechanisms of genome packaging \cite{Chelikani2014}, in many virus families with single-stranded (ss) RNA or DNA genomes, the capsid assembles spontaneously around the genomic NA.
Electrostatics provides an important driving force for assembly around NAs due to the presence of positive charges, located on the inner surface of capsid proteins or on flexible terminal domains that dangle into the capsid interior.

The dynamics of assembly around a NA (or other cargo) has been modeled by Gillespie-type simulations which implicitly build the NA into intermediate states \cite{Dykeman2013a,Dykeman2014} (Fig.~\ref{fig:DykemanFigure}), and with particle-based simulations in which coarse-grained subunits assemble around flexible polyelectrolytes \cite{Perlmutter2013,Perlmutter2014, Perlmutter2015b,Zhang2013,Zhang2013a}, semiflexible polyelectrolytes \cite{Zhang2013a}, or model NAs \cite{Perlmutter2013}. The latter simulations find that assembly around a cargo can proceed through two different mechanisms, see Fig.~\ref{fig:JasonFigure}.

Most viruses with ssRNA or ssDNA genomes are \emph{overcharged}, meaning that the negative charge on the encapsidated RNA significantly exceeds the positive charge on the interior of the capsid \cite{Belyi2006,Hu2008d}. To explain this observation, the length of RNA which optimizes capsid thermostability has been investigated using scaling methods, continuum-models and BD simulations (\eg \cite{Belyi2006,Zandi2009, Erdemci-Tandogan2014,Schoot2013,Perlmutter2013}). These calculations suggest that overcharging arises because only a fraction of RNA charges can reside within the electrostatic screening distance of  capsid charges, and that intra-molecular RNA base-pairing increases overcharging \cite{Perlmutter2013,Erdemci-Tandogan2014,Schoot2013}.

Despite the ability of nonspecific electrostatics to promote assembly around heterologous RNA, viruses package their genomic RNA with remarkable selectivity \emph{in vivo}
(\eg 99\% \cite{Routh2012}). Several factors have been proposed to explain selective packaging. Experiments and simulations suggest that the physical features of viral RNAs (\eg charge, and size due to tertiary structure) are optimized for assembly of their capsid \cite{Tubiana2015,Garmann2015,Perlmutter2013,Singaram2015,Comas-Garcia2012,Erdemci-Tandogan2014,Gopal2014}.
Secondly, interactions between capsid proteins and specific sequences within the genome called packaging sites (PSs) have been identified for a number of viruses (\eg \cite{Rao2006,Stockley2013a,Dykeman2011}). Experiments find that viral genomes contain many PSs (of order 30-60) with binding affinities to capsid proteins ranging from nanomolar to micromolar \cite{Stockley2013a}.

Gillespie-type \cite{Dykeman2013a,Dykeman2014} and BD simulations \cite{Perlmutter2015b} of assembly around RNAs containing PSs predicted preferential packaging over uniform  RNAs (without PSs) for certain parameter ranges, but predicted poor selectivity under conditions which are optimal for assembly around uniform RNAs. However, selectivity was significantly enhanced when considering the steadily increasing concentration of capsid proteins which occurs within a bacterial host (Fig.~\ref{fig:DykemanFigure}) \cite{Dykeman2014}.

Knowledge of the locations of PSs within genomes and models for how PS binding couples to the capsid geometry have been used as constraints for analyzing electron tomography data of RNA within the MS2 capsids \cite{Geraets2015}. The analysis suggested that the encapsidated RNA is highly organized, with similar conformations in most viruses. A tomography study of HBV virions led to a similar conclusion (\eg \cite{Wang2014}).

Additional factors have been proposed to enable selective RNA packaging \textit{in vivo}, including subcellular localization of viral components and coordinated translation and assembly (reviewed in \cite{Rao2014}). These factors have yet to be incorporated into assembly models.

\begin{figure*} [hbt]
\centering{\includegraphics[width=0.9\textwidth]{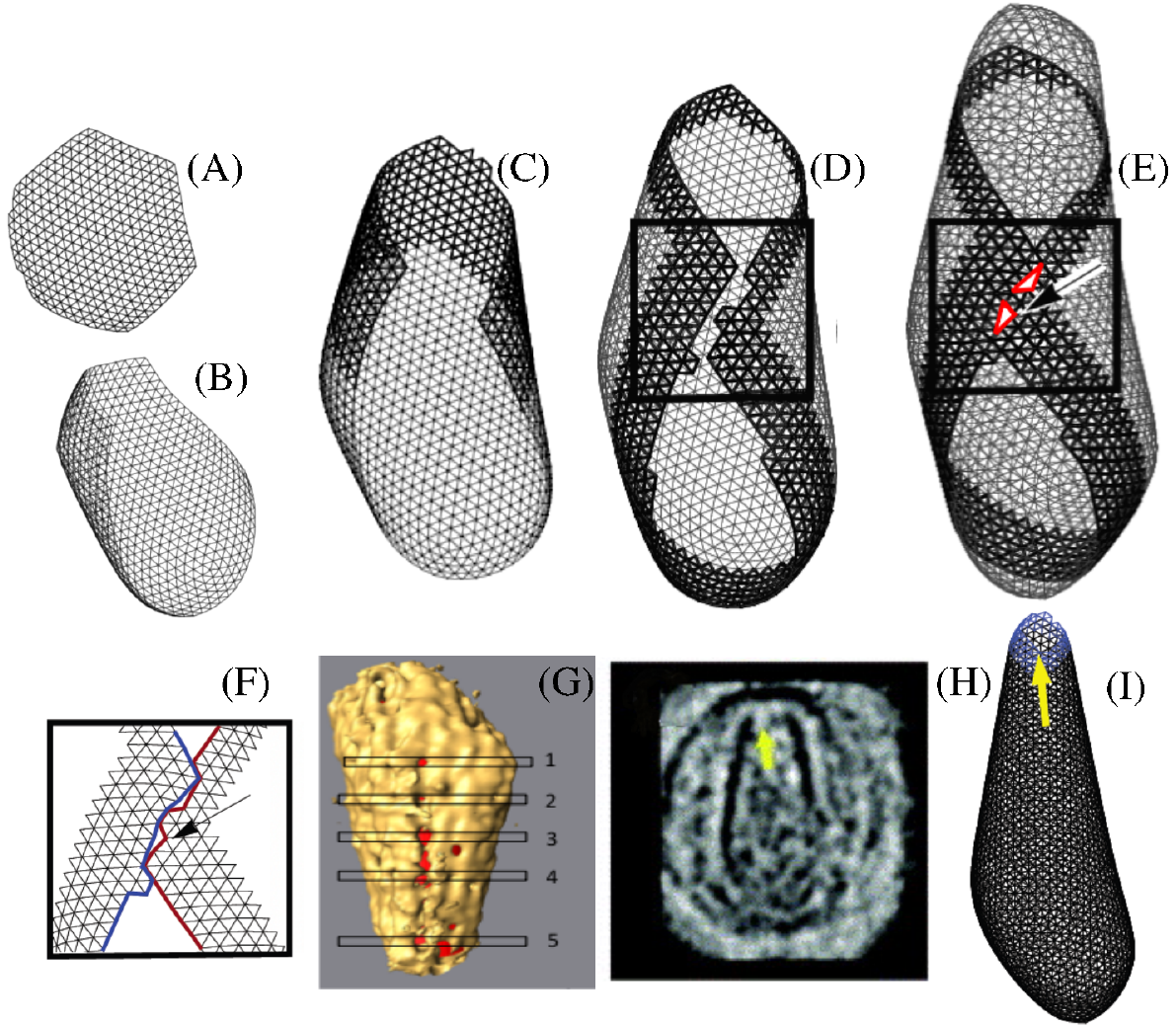}}
\caption{{\bf (A-E)} Snapshots of simulation pathways of a coarse-grained model for HIV mature capsid assembly by the de novo mechanism discussed in the text. {\bf (A)} A triangular lattice grows with local hexagonal symmetry. {\bf (B)} Due to the inherent spontaneous curvature of subunits, a few pentamers form as the shell grows, inducing a region of high curvature. {\bf (C)} The shell continues to grow, and curls over. {\bf (D)} Eventually, two edges become close enough to merge. {\bf (E)} Depending on the positions of subunits at the growing edge at the time of merging, a line of defects (seam) can form, as observed in experiments. {\bf (F)} A close-up of the seam.
{\bf (G-H)} Cryotomographic reconstructions of HIV mature capsids confirm that some capsids are defective, showing {\bf (G)} seams  or {\bf (H)} holes at capsid tips (yellow arrow).  {\bf (I)} A simulated shell in which accumulation of strain at the tip prevents closure, resulting in a hole at the tip, as observed in experiments. All figures are reproduced from Yu et al.\cite{Yu2013}.
}
\label{fig:RoyaFigure}
\end{figure*}

{\it Assembly of HIV.} 
In contrast to the viruses described above, the capsid of the human immunodeficiency virus (HIV) lacks icosahedral symmetry. The virus initially assembles as an `immature' spherical particle, constructed of a disordered lattice of uncleaved Gag proteins and surrounded by a lipid bilayer. The latter is derived from the plasma membrane of the host cell during budding (exiting) of the virus from the cell. Upon maturation, Gag polyproteins are cleaved into three distinct portions, the matrix (MA), capsid (CA), and nucleocapsid (NC) proteins, which constitute different components of the fully infectious virions \cite{Sundquist2012,Ganser-Pornillos2008}. The MA proteins are bound to the interface of the bilayer envelope, forming an outer shell. Inside, about 1500 of the CA proteins assemble into an unusual shell around a condensed complex of the RNA and NC proteins \cite{Sundquist2012}. In HIV-1, the CA shell forms as a ``fullerene cone'': a hexagonal lattice containing 12 pentamers, usually with 5 pentamers at the cone tip and 7 at the base \cite{Ganser1999}. Molecular dynamics flexible fitting (MDFF), a technique in which MD is performed using constraints based on cryo-EM data, recently enabled  atomic-resolution models for structures of the mature HIV capsid \cite{Zhao2013}. All-atom MD simulations on the resulting structure were used to examine contacts between hexamers, pentamers and various assembly units, which showed that the presence of pentamers gives rise to closer trimer contacts and higher surface curvature.

Despite cryo-electron tomographic studies \cite{Briggs2006,Sundquist2012,Woodward2015}, the structural details of the immature particle and the maturation pathway remain unknown. Two models have been proposed for how HIV CA proteins assemble into the mature conical shells. In the ``displacive" model, the CA lattice from the immature shell does not completely disassemble, but undergoes conformational changes to build the mature capsid \cite{Frank2015,Meng2012,Keller2013}. In the other model, the immature spherical shell completely disassembles after Gag cleavage, followed by ``de novo" reassembly of CA into the cone. Even among researchers who favor the de novo model, there is currently no consensus on the order of assembly of the cone.  According to Briggs et al. \cite{Briggs2006}, the capsid nucleates from a narrow tip and then grows until hitting the opposite side of the enclosing membrane, at which point the cone base forms. Alternatively, the experiments of Benjamin {\it et al.} \cite{Benjamin2005} reveal a small hole (defect) on cone tips, suggesting an opposite pathway, in which the cone base assembles first, followed by formation of the body and then the narrow tip.

While there have been no modeling studies corresponding to the displacive model, several coarse-grained models have been employed to study the de novo assembly of conical capsids. Since experiments show that the CA protein can assemble to form conical and cylindrical shells {\it in vitro} in the absence of genome \cite{Ganser1999,Meng2012,Zhao2013}, thus far simulations have focused on assembly of conical shells in the absence of a membrane or genome.  As the curvature of cone varies constantly along its axis of symmetry, a question naturally arises: how do the proteins adjust to sit in very different environments, and what are the asssembly pathways?

The growth of retrovirus shells was studied using the growing elastic sheet model described above \cite{Hicks2006}.  While most of their simulated shells had irregular structures, addition of an attractive interaction between nearby edges of the growing shell (mimicking hydrophobic interactions) led to conical shells for large values of the Foppl von Karman number (meaning that the protein shell ˜resists stretching more strongly than bending) \cite{Levandovsky2009,Yu2013}.  These simulations also explained the `seams' observed in EM structures of mature capsids. Combining the model with experiments suggested that the HIV capsid can start to grow as a hexagonal lattice that eventually forms the body of the shell, with the tip and base of the cone closing toward the end of the assembly (Fig.~\ref{fig:RoyaFigure}) \cite{Yu2013}.

Several particle-based models for CA dimers have been developed, in which subunit shapes and short-ranged interactions between specific sites on subunits were based on solution NMR structures or all-atom MD simulations \cite{Qiao2015,Grime2014,Zhao2013}. Because solution NMR studies identify flexibility between the N-terminal and C-terminal domains within a dimer, simulations included flexible dimers  \cite{Grime2012} or ensembles of dimers with different configurations based on an all-atom MD simulation \cite{Qiao2015}. Simulations of the assembly process (reaching up to 20-40 \% of a complete capsid) found that  associations between trimer-of-dimers intermediates play a key role in the assembly process. Relatedly, fitting of a rate equation model against in vitro experiments in which CA proteins assemble into tubes identified a trimer-of-dimers as the critical nucleus \cite{Tsiang2012}.

\section*{Outlook}
The coarse-grained models presented in this review have elucidated key aspects of the virus formation process, which are not accessible to experiments or all-atom simulations. Looking ahead, an important area which is only beginning to be addressed by coarse-grained models is how the environment of a host cell contributes to viral assembly. In conjunction with quantitative experiments studying the effects of specific host factors, such models can advance our understanding of how viruses propagate, and pave the way for discovering new approaches to combat viral infections.

\section*{Acknowledgements}
The acknowledge support from NIH grant R01GM108021 (MFH), the  NSF Brandeis MRSEC,  DMR-1420382 (MFH) and  NSF grant DMR-1310687 (RZ).

\end{document}